\documentclass{iaus}
\usepackage{graphicx}

\title[Doppler Boosting and Deboosting Effects] 
{Doppler Boosting and Deboosting Effects in the Forward Relativistic Jets of
AGNs and GRBs}

\author[Zhou and Su] 
{Jianfeng Zhou$^1$,
 Yan Su$^2$}

\affiliation{$^1$Department of Engineering Physics, Center for Astrophysics, Tsinghua University,
Beijing, 100084, China \break email: zhoujf@tsinghua.edu.cn\\[\affilskip]
$^2$National Astronomical Observatories, Chinese Academy of Sciences,
    Chaoyang District, Datun Road 20A, Beijing, 100012, China 
   \break email: suyan@bao.ac.cn}

\pubyear{2006}
\volume{238}  %% insert here IAU Highlights of Astronomy Volume Number
\pagerange{001--999}
\date{?? and in revised form ??}
\jname{Black Holes: from Stars to Galaxies -- across the Range of Masses}
\editors{V. Karas \& G. Matt, eds.}
\begin{document}

\maketitle

\begin{abstract}
It is widely accepted that the Doppler deboosting effects exist in counter relativistic
jets. However, people often neglect another important fact that both Doppler
boosting and deboosting effects could happen in forward relativistic jets. Such effects
might be used to explain some strange phenomena, such as the invisible gaps
between the inner and outer jets of AGNs, and the rapid initial decays and re-brightening
bumps in the light curves of GRBs.
\end{abstract}

\firstsection % if your document starts with a section,

\section{Doppler factors in relativistic jets}
In the relativistic jets of AGNs or GRBs, the
observed flux is related to their intrinsic flux by
$F_{\rm obs}=\delta^{3+\alpha} F$, where $\delta$ is Doppler factor,
$F_{\rm obs}$ and $F$ are the observed and intrinsic flux respectively , and $\alpha$ is
the spectral index (\cite[Blandford \& K\"{o}nigl 1979]{bla79}). If $\delta$ is greater 
than 1, then the observed
flux will be enhanced, which is called
Doppler boosting effect. On the other hand,
if $delta$ is less than 1, the observed flux is attenuated,
which is named to Doppler deboosting effect.

The Doppler factor of a jet can be described by the following equation, 
\begin{equation} \label{eqn1}
\delta = \frac{1}{\gamma(1-\beta\cos\theta)} 
\end{equation}
where $\beta$ is the velocity, $\gamma$ is the Lorentz factor, and $0 \leq \theta \leq \pi$
is the viewing angle.

\begin{figure}
\center{\includegraphics[height=5cm,width=5cm]{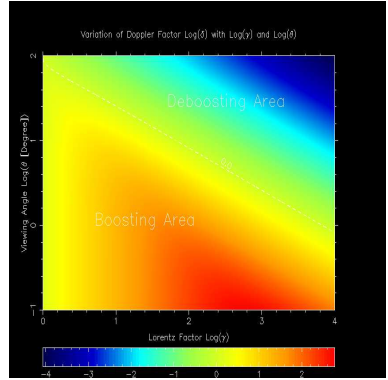}
\caption{The function of Doppler factor in terms of viewing angle and Lorentz
factor. Doppler factor, Lorenz factor and viewing angle are all in logarithmic
scale. A contour line, corresponding to $\log\delta = 0$, is plotted. This line cuts the
figure into two areas, i.e. Doppler boosting area and deboosting area.}} \label{fig01}
\end{figure}

For a counter relativistic jet where $\theta > \pi/2$, the Doppler factor $\delta$ is always less 
than 1, which can
easily be derived from equation~\ref{eqn1}. So a counter jet is always Doppler deboosted. For a
forward relativistic jet, the situation is more complex. Provided that the viewing angle
of jet $\theta > 0$ (when $\theta=0$, $\delta$ is always greater that 1), the Doppler 
factor can be greater
than 1 as well as less than 1 (see figure~\ref{fig01} for detail). Therefore, both Doppler boosting
and deboosting effects could happen in forward jets.

\section{Application to AGNs and GRBs}
\subsection{Some observational facts}
In many radio loud AGNs, the large scale jets share some common features in their profiles.
Firstly, there are compact and bright cores in the
center of jets. Secondly, adjacent to the cores, the flux of the jets
drops down very quickly, even form some gaps where
the jets are undetectable. Thirdly, the jets will be re-brightened in the outer
region.

Recently, the SWIFT found some interesting properties in GRBs. Five GRBs'
X-ray light curves are characterized by a rapid fall-off for first few hundred
seconds, followed by a less rapid decline lasting several hours. The light curves 
also show discontinuity (\cite[Tagliaferri et al. (2005)]{tag05}). \cite{bur05}
found that there were bright X-ray flares in GRB afterglows.

\subsection{Explanation}
To qualitatively explain the profiles of the jets in AGNs as well as the 
light curves in GRBs, we divide the evolution of the jets into four stages :
\begin{itemize}
\item Stage I : The jets are accelerating and boosting, which relates to 
the bright cores of AGNs and the bursts of GRBs. The initial acceleration of relativistic jets has been 
detected in 3C 273 by \cite{kri01} and modeled by \cite{zjf04}. In this stage, $\gamma$ is usually
less than a few tens, and $\delta$ increase very quickly.
\item Stage II : The jets are accelerating but deboosting. As the acceleration continues,
 $\gamma$ will be very large ($>100$). Therefore, the jets will enter into
the deboosting area, i.e. $\delta<1$. In this stage, the observed flux of the jets decreases
very quickly, and often forms the gaps between inner and outer jets in AGNs and the rapid decays and
discontinuous light curves in GRBs.
\item Stage III : The jets are decelerating and boosting. The acceleration, however, won't last
forever because of radiation loss and the interaction between the jets and the surrounding medium.
Thus, the jets will decelerate and their $\delta$ will increase again. Consequently, the
Doppler boosted jets will appear again in the profiles of AGNs or in the light curves of GRBs. 
\item Stage IV : The jets are decelerating and deboosting. Finally, due to the same reasons in 
stage III, the jets will gradually disappear.

\end{itemize}

\end{document}